\documentclass[journal]{IEEEtran}
\usepackage{graphicx}
\usepackage{epstopdf}
\usepackage{amssymb}
\usepackage{afterpage}

\usepackage{color}
\graphicspath{{figures/}}
\usepackage{tabularx}
\usepackage{multirow}
\usepackage{booktabs}
\usepackage{enumitem}
\usepackage{flushend}
\usepackage{amsmath}

\begin{document}

\title{A deep learning approach for virtual monochromatic spectral CT imaging with a standard single energy CT scanner}
\author{Wei~Zhao$^\dag$,~
        Tianling~Lyu$^\dag$,~
        Yang~Chen$^*$,~
        and~Lei~Xing$^*$~
\thanks{We gratefully acknowledge the support of NVIDIA Corporation
for the GPU donation. \emph{W. Zhao and T. Lyu contributed equally to this work and asterisk indicates corresponding authors.}}
\thanks{W. Zhao is with the Department of Radiation Oncology, Stanford University, Stanford, CA, 94305 USA.}
\thanks{T. Lyu is with the Department of Radiation Oncology, Stanford University, Stanford, CA, 94305 USA and also with the Department of Computer Science and Engineering, Southeast University, Nanjing, Jiangsu, China.}
\thanks{Y. Chen is with the Department of Computer Science and Engineering, Southeast University, Nanjing, Jiangsu, China (e-mail: chenyang.list@seu.edu.cn).}
\thanks{L. Xing is with the Department of Radiation Oncology, Stanford University, Stanford, CA, 94305 USA (e-mail: lei@stanford.edu).}
}

%



\maketitle

\begin{abstract}

Purpose/Objectives: To develop and assess a strategy of using deep learning (DL) to generate virtual monochromatic CT (VMCT) images from a single-energy CT (SECT) scan.
Materials/Methods: The proposed data-driven VMCT imaging consists of two steps: (i) using a supervised DL model trained with a large number of 100 kV and 140 kV dual-energy CT (DECT) image pairs to produce the corresponding high-energy CT image from a low-energy image; and (ii)  reconstructing VMCT images with energy ranging from 40 to 150 keV. To evaluate the performance of the method, we retrospectively studied 6,767 abdominal DECT images. The VMCT images reconstructed using both DL-derived DECT (DL-DECT) images and the images from DECT scanner were compared quantitatively. Paired-sample $\textbf{\emph{t}}$-tests were used for statistical analysis to show the consistency and precision of calculated HU values.
Results: Excellent agreement was found between the DL-DECT and the ground truth DECT images ($\textbf{\emph{p}}$ values ranged from 0.50 to 0.95). Noise reduction up to 68\% (from 163 HU to 51 HU) was achieved for DL-based VMCT imaging as compared to that obtained by using the standard DECT. For the DL-based VMCT, the maximum iodine contrast-to-noise ratio (CNR) for each patient (ranging from 15.1 to 16.6) was achieved at 40 keV. In addition to the enormous benefit of VMCT acquisition with merely a SECT image, an improvement of CNR as high as 55\% (from 10.7 to 16.6) was attained with the proposed approach.
Conclusions: This study demonstrates that high-quality VMCT images can be obtained with only a SECT scan.

\end{abstract}



%
\IEEEpeerreviewmaketitle

\section{Introduction}

\IEEEPARstart{W}{ith} the development of dual-energy CT (DECT) data acquisition and reconstructed techniques~\cite{alvarez1976energy,kalender1986evaluation,flohr2006first,boll2008calcified,mccollough2015dual,zhao2016using,zhao2018unified}, virtual monochromatic CT (VMCT) imaging from DECT with two polychromatic x-ray measurements has gained increasing popularity for its unique capability in mimicking CT images with monochromatic x-ray photons~\cite{goodsitt2011accuracies,yu2011virtual}. By inheriting the attenuation properties of the monochromatic beam, VMCT imaging can mitigate the beam-hardening artifacts and energy-shift phenomena widely existing in conventional polychromatic CT images. With the increased consistency and precision of CT numbers, the VMCT imaging enables us to interrogate the changes in attenuation over a range of diagnostic energy levels. Hence, VMCT imaging has shown improvement in image quality compared with conventional polychromatic CT and is valuable for diagnosis of various diseases~\cite{matsumoto2011virtual,marin2014dual,yu2012dual,pomerantz2013virtual}. The modality is also useful to reduce metal artifacts~\cite{pessis2013virtual}, overcome renal cyst pseudoenhancement~\cite{mileto2014impact}, and increase the visibility of low-contrast lesions~\cite{leng2015maximizing}. For these reasons, there is growing interest in VMCT imaging as a substitute for conventional standard protocol polychromatic CT.

Current VMCT imaging relies on the high-end DECT scanner and its practical implementation for routine application is challenging. Depending on the implementation of DECT imaging, VMCT images can be obtained either in projection-domain or image-domain~\cite{mileto2016virtual}. The projection-domain method has potential to completely eliminate beam-hardening artifacts, while the image-domain is advantageous in providing images acquired using different filtered spectra which results in a better spectra separation. For both methods, the optimal energy level depends on patient size, dual-energy spectra, dose partitioning, and the image quality metric to be optimized~\cite{yu2011virtual,albrecht2016advanced}. In this study, we focus on image-domain VMCT imaging.

Deep learning (DL) uses deep neural network to learn complex relationship from big data and to incorporate prior knowledge into an inference model~\cite{zhao2019markerless,shen2019patient,zhao2019incorporating}. The technique has been used to different applications related to DECT, such as material decomposition~\cite{liao2018pseudo,xu2018projection,zhang2019image,poirot2019physics}, DECT image generation from a SECT image~\cite{li2017pseudo,zhao2019dual,zhao2019deep}, and VMCT imaging from standard DECT~\cite{cong2017monochromatic,feng2018fully}. Here we investigate a DL strategy of using a single SECT image dataset for virtual monochromatic spectral CT imaging. The approach mitigates the need for a premium DECT scanner, which is much less accessible, especially in underdeveloped regions, and significantly improves the contrast-to-noise ratio (CNR) of VMCT imaging. It may thus pave the way for future widespread adoption of VMCT in clinical practice.

\begin{figure*}[t]
    \centering
    \includegraphics[width=0.9\textwidth]{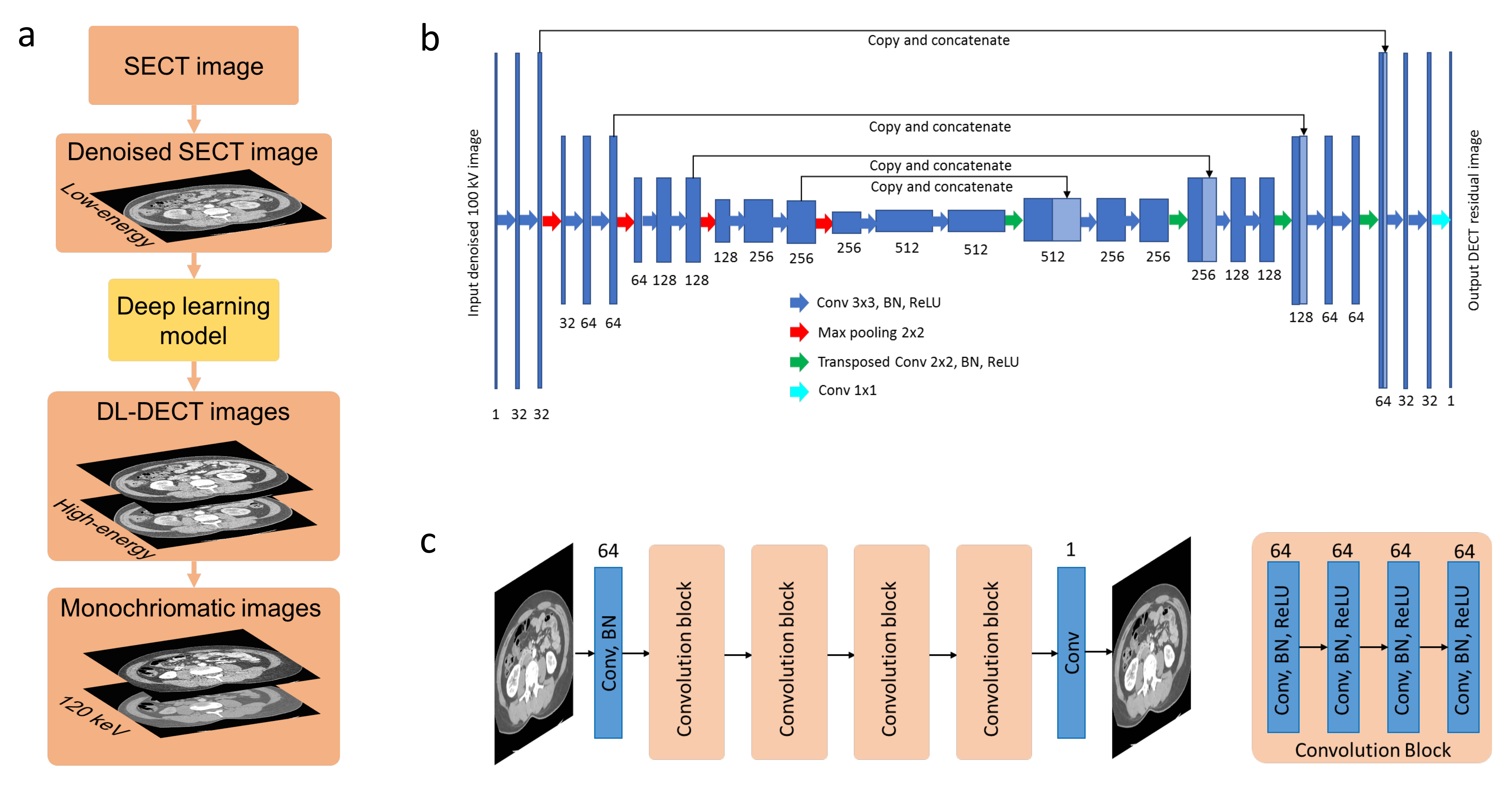}
    \caption{The proposed DL-based VMCT imaging method. a), The overall workflow. b), The architecture of the dual-energy residual mapping network. c), The architecture of the fully convolutional network for image denoising. DL = deep learning, SECT = single-energy CT, DECT = dual-energy CT.}
    \vspace{-1em}
    \label{fig:1}
\end{figure*}
\section{Material and methods}
This retrospective study was approved by the institutional review board and compliant with the Health Insurance Portability and Accountability Act.  The overall workflow for DL-based VMCT imaging is shown in Fig. 1a. Two key tasks of the process include: (i) using a DL model trained with a large number of 100 kV (low energy CT) and 140 kV (high-energy CT) DECT image pairs to produce the corresponding high-energy CT image of 140 kV from an image with an energy of 100 kV; and (ii)  reconstructing VMCT images with energy ranging from 40 to 150 keV using the DL-derived DECT images (DL-DECT). These are described in details in the following along with our evaluation procedure of the VMCT imaging approach.

\subsection{DL-based DECT imaging}

A DL model (Fig. 1b) was developed for deriving high-energy CT images from the low-energy images acquired using a standard SECT scanner. To obtain accurate high-energy DL-DECT images, the model was designed in such way that it takes denoised low-energy image as input and outputs the residual image defined as the difference between the high-energy image and input low-energy CT image (details in section~\ref{dualmapping}).  With the model, the desired high-energy image is derived by adding the output residual image and the input low-energy image together.
For the training of DL-DECT model, we used retrospectively 4,736 pairs of low- and high-energy images of 16 patients (median age: 53 with range from 42 to 78) acquired by using a SOMATOM Definition Flash dual-source DECT scanner (Siemens Healthineers, Forchheim, Germany). The low- and high-energy DECT scans were performed using 100 kV and 140 kV x-ray beams, respectively, with the 140 kV spectrum pre-hardened using a tin filter to enhance dose efficiency and spectral separation. All patients were administrated 300mg I/ml iodine contrast agent and DECT scans were performed 25 seconds later after the contrast agent injection. CT images were reconstructed using a commercial filtered back-projection algorithm with the D30f convolutional kernel which is a medium-smooth body kernel appropriate for dual-energy processing. Before being fed into the DL model, both low- and high-energy DECT images were denoised by using an in-house fully convolutional network (FCN) (Fig. 1c, and details in section~\ref{denoising}). Next, we calculated the residual image between the denoised low- and high-energy CT images and trained a dual-energy mapping network by using the low-energy CT and the residual images as the network input and output.
The validation and testing datasets encompass 1,071 (3 patients, median age: 49 with range from 32 to 66) and 960 CT (another 3 patients, median age: 48 with range from 38 to 49) DECT image slices, respectively. In the validation and testing process, the trained network took a low-energy image as input and generated the corresponding residual image as output, which was then added to the input low-energy image to yield the corresponding high-energy image. The predicted high-energy image and the input low image are denoted as DL-DECT image.

\subsection{Dual-energy mapping}
\label{dualmapping}

A U-Net-type deep neural network was used to generate the residual image from denoised low-energy image. The residual image was then added to the original low-energy image to yield the final high-energy image. The network used an ¡°encoder-decoder¡± architecture with skip connections to learn the residual between the DECT images in an end-to-end fashion. The encoder part includes five convolutional blocks, and each of the blocks encompasses two consecutive 2D convolutional layers, followed by a ReLU layer and a BN layer. Each of the first four blocks is followed by a max-pooling layer which down-samples the spatial information of the features by a factor of two. Meanwhile, the channels of the feature hierarchies were doubled during the blocks by doubling the number of the convolutional filters in each block. The decoder part includes four convolutional blocks, each followed by a transposed convolutional operation which up-samples the features by a factor of two using a fractionally-strided convolution. Before going through the convolutional blocks, the input of each blocks in the decode part was concatenated with the corresponding features from the encoder part. The output of the decoder was fed into a convolution layer with $1\times1$ to reduce the dimensionality in the filter dimension, resulting a feature activation that has the same size as the desired dual-energy difference image. During training, we used the mean-squared error loss to minimize the difference between the final activation map and the ground truth image, which is the difference image between the denoised 100 kV and 140 kV images.

\subsection{Image denoising using Fully Convolutional Network (FCN)}
\label{denoising}
The FCN model was constructed using an input layer and 16 consecutive convolution blocks followed by an output layer. The input and output layers perform convolution operations using 64 kernels with size of $3\times3$ and 1 kernel with size of $3\times3\times64$, respectively. Both convolution operations use stride $1\times1$ and padding 1. Each of the 16 convolution blocks consists of a convolutional layer, followed by a batch normalization (BN) layer for network stability improvement and convergence acceleration, and a rectified linear unit activation (ReLU) layer. For each of the convolutional layer, 64 convolution kernels with size of size $3\times3\times64$, stride $1¡Á\times1$, padding 1 were used and the features maps remain the same through all the convolution blocks. The final activation map of the FCN has the same size as the input image and was compared to the difference image using a weighted L2 loss function. During network training procedure, the networks weights were iteratively updated by backpropagating the residual of the loss function.
We retrospectively used publicly accessible patient data from American Association of Physicists in Medicine (AAPM) Low-Dose CT Grand Challenge. Ten patient cases with a range of sizes were included. For each case, full-dose and quarter-dose CT images were provided. To train the denoising network, we regarded the quarter-dose images as orginal CT images and the full-dose images as noise reduced CT images. With this, the quarter-dose image and its residual image with respect to the full-dose image were used as the model input and output, respectively. Since the FCN is independent of the input image size, we divide the original $512\times512$ DICOM CT images into $64\times64$ image patches to further increase the number of training samples during the training phase.
Once the model was trained, we deployed it to the DECT data for both low- and high-energy CT images. In this inference procedure, the model output (i.e. the residual image) were then subtracted from the input original DECT images to yield noise reduced DECT images.

\begin{figure*}
    \centering
    \includegraphics[width=0.85\textwidth]{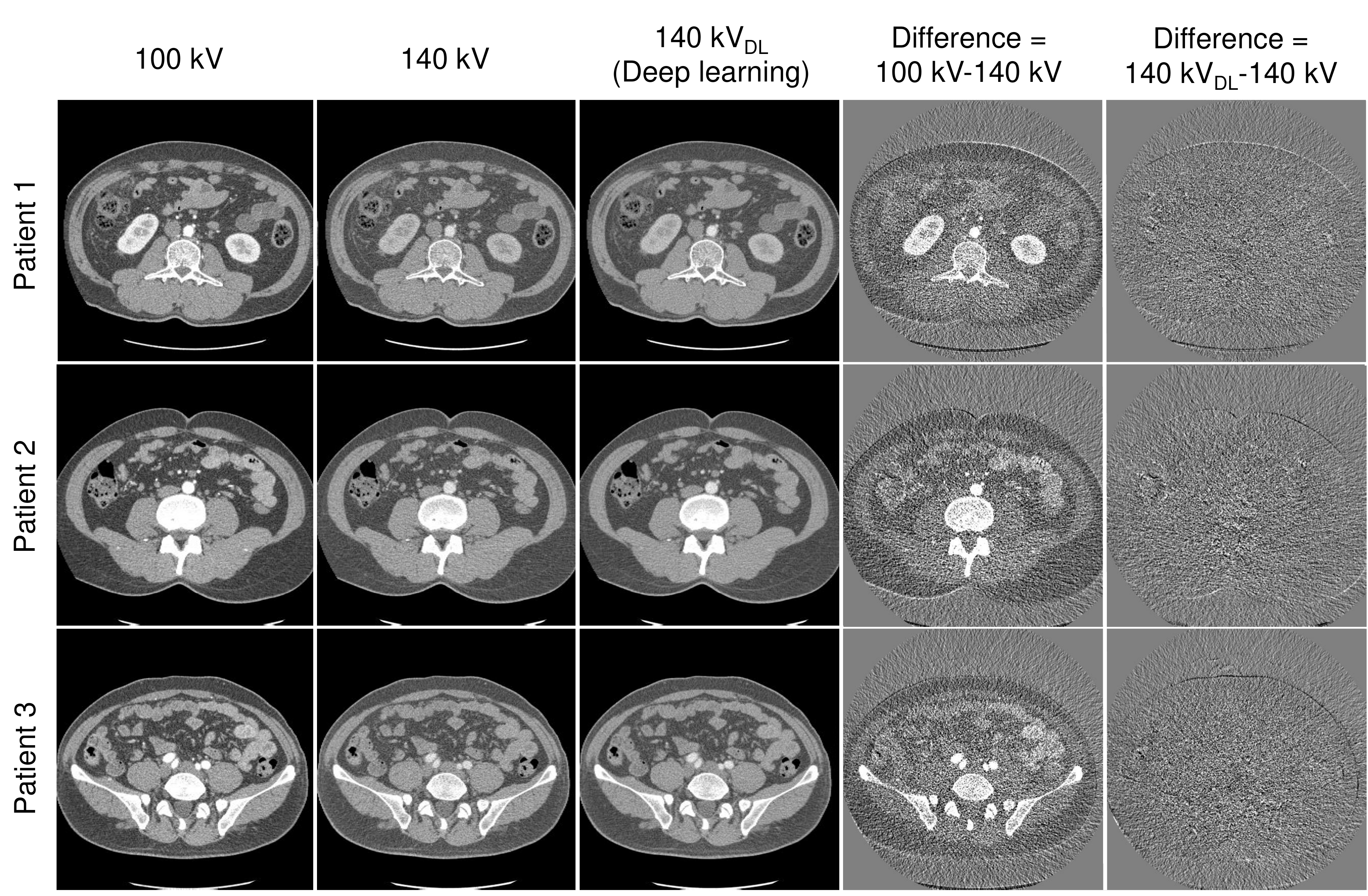}
    \caption{Original 100kV/140kV DECT images and DL-predicted 140kV images and their difference images with respect to the ground truth images for three testing patients who underwent contrast-enhanced DECT scans. The first and second columns show the original 100 kV and 140 kV images. The third and fifth columns show the DL-predicted 140 kV images and their differences with respect to the original 140 kV images. For comparison, the fourth column shows the difference images between the original DECT images. All CT images are displayed with C=0 HU/W = 500 HU, while the difference images are displayed in a tighter window (C=0 HU/W = 200 HU). DL = deep learning, DECT = dual-energy CT.}
    \vspace{-1em}
    \label{fig:2}
\end{figure*}

\begin{table*}
\centering
\caption{Quantitative comparisons of HU accuracy of the original 140 images, the FCN denoised 140 kV images, and the DL-predicted 140 kV images.}
\resizebox{0.75\textwidth}{!}{\begin{tabular}{lcccccc}
\toprule
\multicolumn{2}{ c }{\multirow{2}{*}{Patient ID}} & \multirow{2}{*}{Original$^\ast$} &\multirow{2}{*}{Denoised$^\ast$} &\multirow{2}{*}{DL$^\ast$} &\multicolumn{2}{ c }{P Value$^\dag$}\\ \cline{6-7}
 &&&& & Original vs Denoised & Original vs DL \\
\midrule
\multirow{2}{*}{1} & Iodine &178.3$\pm$31.4 &	179.8$\pm$13.1 & 178.5$\pm$24.6 &0.46 &0.95\\
& Tissue &51.1$\pm$34.2 &	51.7$\pm$8.7 & 53.7$\pm$24.3 &0.80 &0.54\\
\midrule
\multirow{2}{*}{2} & Iodine &184.4$\pm$28.7 &	185.3$\pm$6.7 & 185.9$\pm$24.4 &0.68 &0.68\\
& Tissue &51.7$\pm$27.6 &	52.2$\pm$7.0 & 53.8$\pm$20.4 &0.79 &0.50\\
\midrule
\multirow{2}{*}{3} & Iodine &186.2$\pm$35.8 &	186.6$\pm$7.7 & 184.6$\pm$28.9 &0.88 &0.69\\
& Tissue &54.4$\pm$30.8 &	55.5$\pm$7.0 & 56.3$\pm$22.2 &0.61 &0.56\\
\midrule
\bottomrule
\vspace{0.1em}
\end{tabular}}\\
\label{tab:1}
\footnotesize{DL = Deep learning, DECT = dual-energy CT, BM = bone marrow, KN = kidney.\\
$^\ast$ Data are mean $\pm$ standard deviations.\\
$^\dag$ $P<.05$ is defined as the significance level.\\}
\end{table*}

\subsection{Network training}

The FCN and the dual-energy mapping network were trained using a GPU workstation which equipped with 5 Nvidia Titan X GPUs. All weights in the convolution kernels of the networks were initialized using random variables (mean value = 0, variance = $10^{-3}$) with Gaussian distribution. We used the adaptive moment estimation (ADAM) algorithm to optimize the loss functions. The deep learning framework Tensorflow (version r1.9) was used to implement the dual-energy mapping network, which was trained using 200 epochs. The learning rates were set to $10^{-3}$ for the first 150 epochs and $10^{-4}$ for the rest 50 epochs. The network training took about 24 hours.

\subsection{Virtual Monochromatic Imaging}

The DL-DECT images were employed to reconstruct VMCT images of energy from 40 to 150 keV with 10-keV intervals. To this end, we used an image-based method [7] to create VMCT image $\mu(E)$ at energy $E$, which was basically a linear combination of the low-energy image $\mu_L$ and high-energy CT image $\mu_H$, i.e.,
\begin{equation}\label{equ:mono}
\mu(E)=w(E) \mu^L+(1-w(E))\mu^H,
\end{equation}
The energy dependent weight $w(E)$ was calculated using the energy-dependent linear attenuation coefficients of the basis materials and attenuation coefficients $\mu_i^j  (j=L,H;i=1,2)$  of the basis materials in the low- and high-energy images. In this study, water and cortical bone were employed as basis materials and $\mu_i^j$ was calculated using region-of-interests (ROIs) placed on the two materials.
A square $11\times11$ pixel ROI was placed at the aorta and the mean CT number was recorded as the iodine contrast $HU_{iodine}$(shown in Figure 3). A second square ROI of the same size was placed in the background tissue region close to the aorta (shown in Figure 3) and CT number $HU_{tissue}$ and noise $N_{tissue}$ were measured. Iodine $CNR$ was calculated as follows:
\begin{equation}\label{equ:mono}
CNR=(HU_{iodine}-HU_{tissue})/N_{tissue}.
\end{equation}
CNR was calculated for all energy levels of the VMCT images generated using both original DECT (measurement acquired from DECT scanner) and DL-DECT images.

\subsection{Image Analysis and Evaluation}

The DL-derived high-energy images and the VMCT images reconstructed by using the DL-DECT images were compared quantitatively with the original DECT and the corresponding VMCT images, respectively, for all the 3 testing cases. Pixelwise error were obtained by calculating the HU difference value in all the image-to-image comparisons. ROI analysis was performed on the iodine-enhanced aorta and tissue background to compute the HU error between the DL-DECT and original DECT. In order to show the consistency and accuracy of the calculated HU value, paired-sample $t$ tests were performed from pairwise comparison between (i) the denoising DECT images and original DECT images; (ii) the DL-predicted high-energy CT images and original high-energy CT images; and (iii) the VMCT images reconstructed from DL-DECT and original DECT images.


\section{RESULTS}

\begin{figure*}[t]
    \centering
    \includegraphics[width=0.9\textwidth]{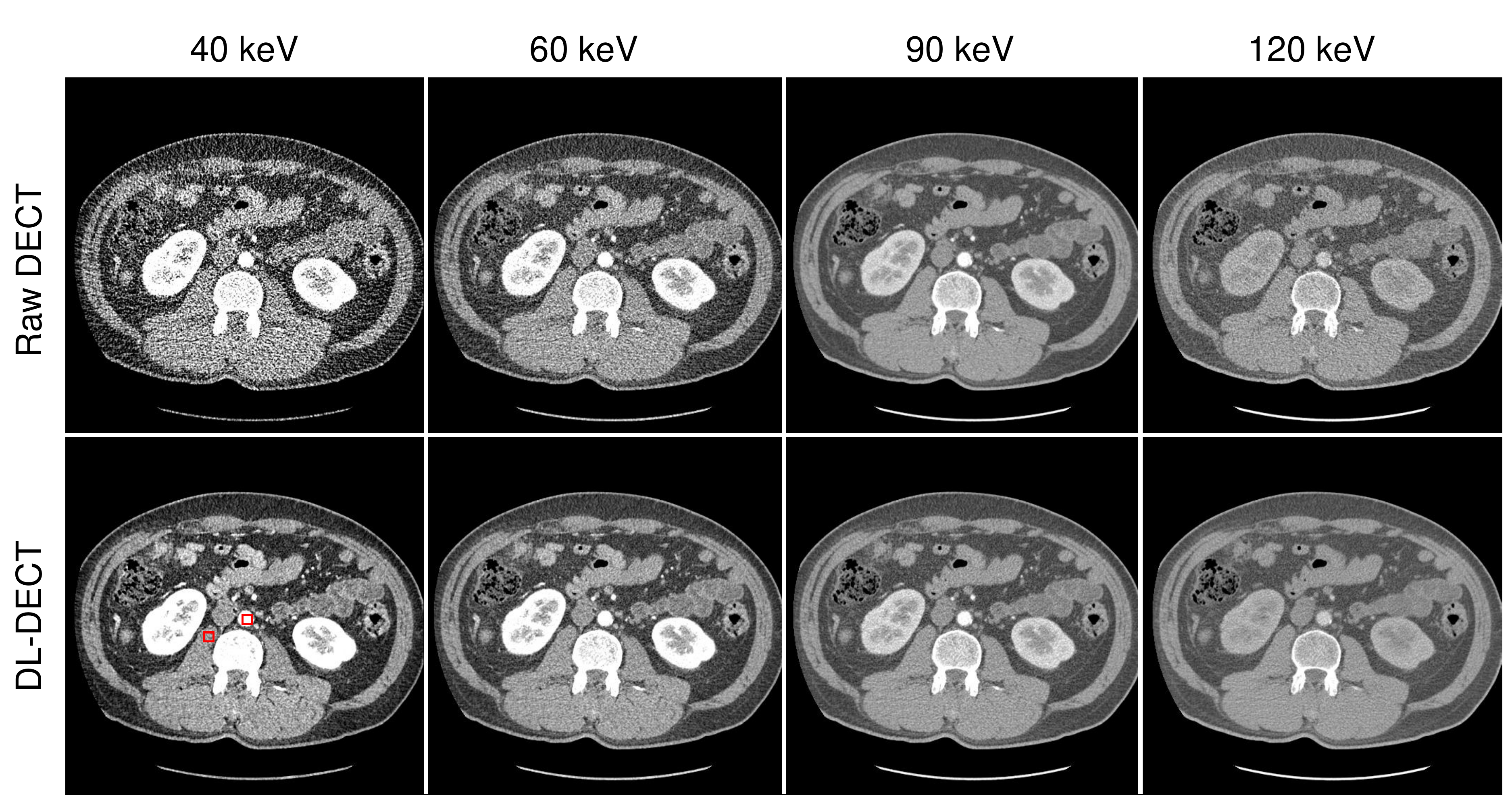}
    \caption{Original 100kV/140kV DECT images and DL-predicted 140kV images and their difference images with respect to the ground truth images for three testing patients who underwent contrast-enhanced DECT scans. The first and second columns show the original 100 kV and 140 kV images. The third and fifth columns show the DL-predicted 140 kV images and their differences with respect to the original 140 kV images. For comparison, the fourth column shows the difference images between the original DECT images. All CT images are displayed with C=0 HU/W = 500 HU, while the difference images are displayed in a tighter window (C=0 HU/W = 200 HU). DL = deep learning, DECT = dual-energy CT.}
    \vspace{-1em}
    \label{fig:3}
\end{figure*}

\subsection{DL-DECT imaging}

The original and DL-predicted 140 kV images are shown in Figure 2 for three testing patients who underwent contrast-enhanced DECT scans. The first and second columns show the original 100 kV and 140 kV images, respectively. The third and fifth columns show the DL-predicted 140 kV images and their differences with respect to the original 140 kV images, respectively. For comparison, difference images between the original DECT images are also shown in the fourth column. As can be seen, the DL-predicted 140 kV images are highly consistent with their corresponding ground truth images, especially for organs and tissues where there are large HU differences between the original DECT images (such as aorta, bone and kidney, shown in the fourth column). Meanwhile, except some insignificant HU differences at the anatomical boundaries, little anatomical structures in the fifth columns are seen, which is attributed to the inherent anatomical differences between the original 100 kV and 140 kV images. These results suggest that the anatomical information is well preserved in the DL-predicted 140 kV images. Quantitative HU accuracy measurements are shown in Table~\ref{tab:1} and there is no significant difference between the predicted and the original 140 kV images (p values ranged from 0.50 to 0.95). For comparison, the HU accuracy of the denoised 140kV images were also included.

\begin{figure}[t]
    \centering
    \includegraphics[width=\linewidth]{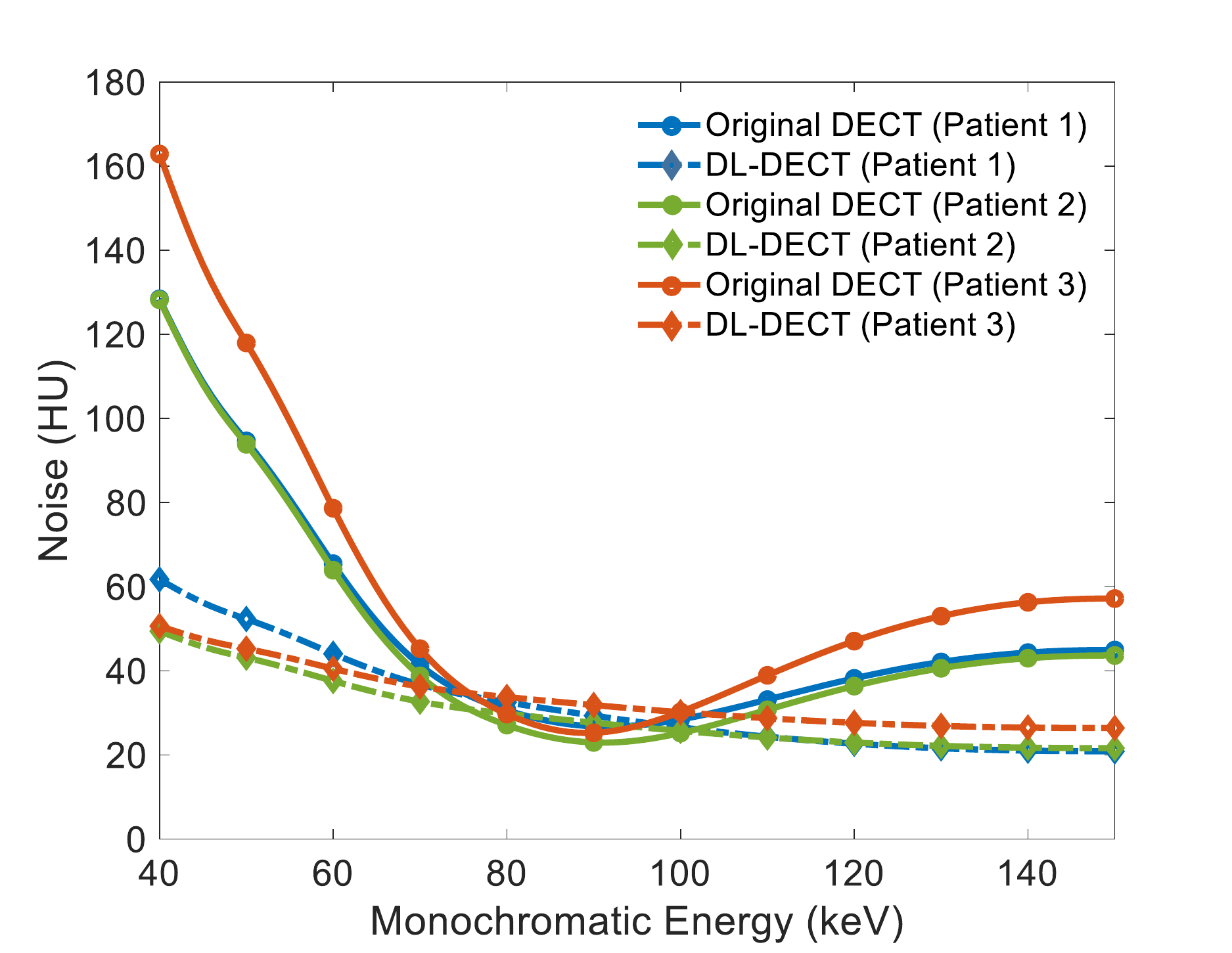}
    \caption{Image noise of virtual monochromatic CT images reconstructed using original DECT and DL-DECT images for three patients at energies ranging from 40 keV to 150 keV. DL = deep learning, DECT = dual-energy CT.}
    \vspace{-1em}
    \label{fig:4}
\end{figure}

\subsection{VMCT imaging}
VMCT images reconstructed using original DECT images and DL-DECT images at 40, 60, 90, 120 keV are shown in Figure 3. The iodine contrast and noise vary a lot in the images reconstructed using original DECT images. For the images reconstructed using DL-DECT images, the iodine contrasts are consistent with that in the images reconstructed using original DECT images (p values are 0.95, 0.68, and 0.69 for the three testing patients, respectively), but the noise levels are relatively stable and much lower than that in the images reconstructed using original DECT images at both low energy (40 keV) and high energy (120 keV).

The relationship between image noise and the energy level of the VMCT images is shown in Figure 4 for the three testing cases both original DECT and DL-DECT scenarios. For the monochromatic images reconstructed from original DECT images, the noise rapidly decreases from 40 keV and reaches a minimum around 90 keV and then increases slowly from 90 to 150 keV. For the DL-based VMCT images, the noise decreases slowly as energy increases from 40 to 150 keV and the noise level is much smaller than that in the VMCT images reconstructed from original DECT except in the 75 to 100 keV energy range, where the DL-based VMCT has slightly higher noise. At 40 keV, the noise reduction of DL-based VMCT is 52\% (from 128 HU to 62 HU), 61\% (from 128 HU to 49 HU), and 68\% (from 163 HU to 51 HU) for patient 1, 2, and 3, respectively. At 140 keV, the noise reduction is 52\% (from 44 HU to 21 HU), 49\% (from 43 HU to 22 HU), and 53\% (from 56 HU to 26 HU) for patient 1, 2, and 3, respectively. At 90 keV, the noises of VMCT images from the original DECT and DL-DECT are found to be (27 and 29 HU), (23 and 28 HU), and (25 and 32 HU) for patient 1, 2, and 3, respectively.
The iodine CNRs as a function of energy for both VMCTs from original DECT and DL-DECT are shown in Figure 5 for the three patients. In the former scenario, iodine CNR starts increasing from 40 keV and reaches to the maximum at around 80-85 keV. It then decreases with the energy. In the latter case, iodine CNR exhibits a very different behavior. It decreases continuously as the energy is changed from 40 keV to 150 keV. Overall, the iodine CNR of DL-based VMCT is much higher than that of the original DECT scenario with only a minor degradation at around 80-90 keV. Table~\ref{tab:2} provides a comparison of the iodine CNR of the original 100 kV images and the maximum achievable iodine CNR for the two different types of VMCT images. For all patients, the maximum iodine CNR is achieved by using the VMCT images reconstructed using DL-DECT.

\begin{figure}[t]
    \centering
    \includegraphics[width=\linewidth]{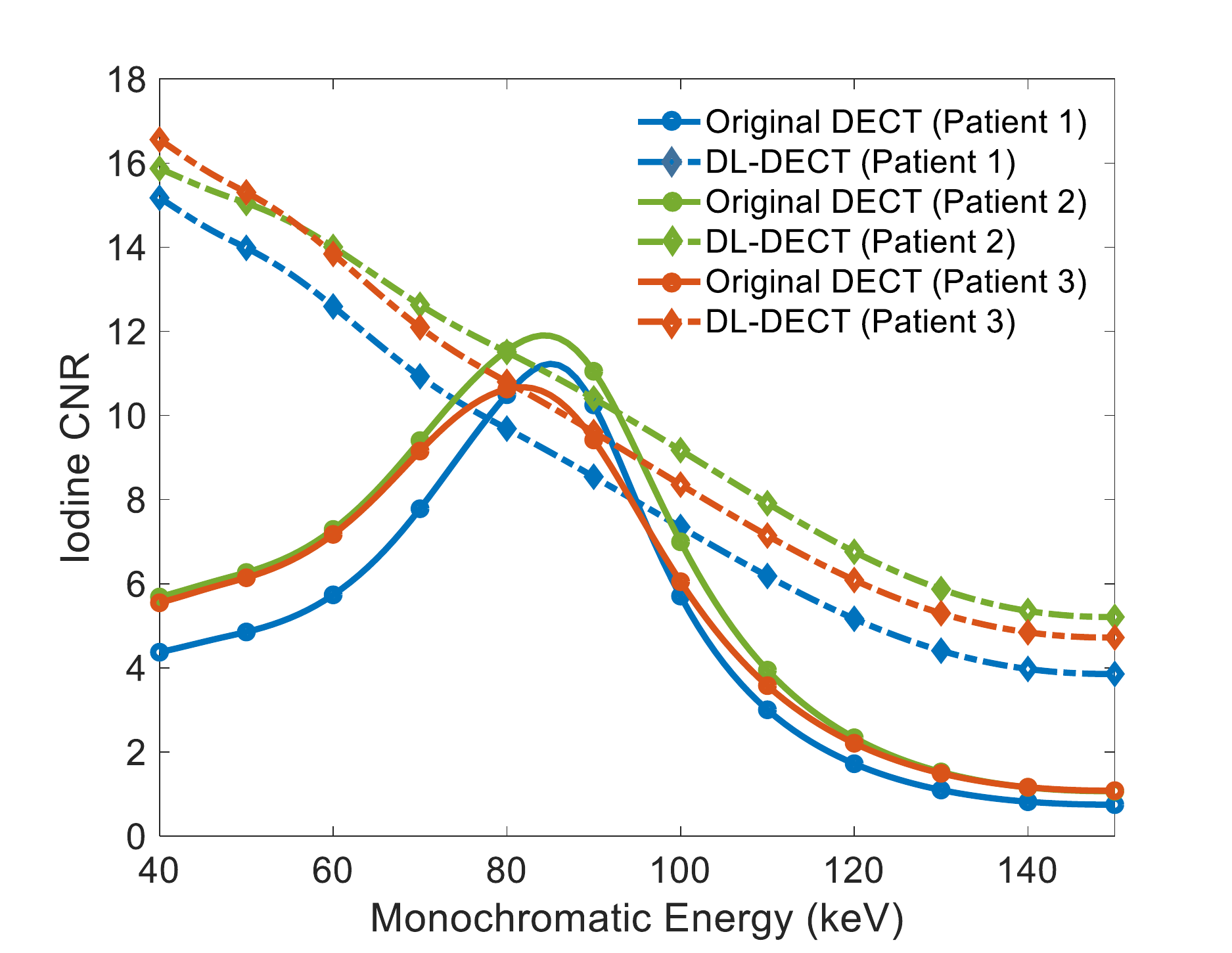}
    \caption{Iodine CNR of virtual monochromatic CT images reconstructed using original DECT and DL-DECT images for three patients at energies ranging from 40 keV to 150 keV. CNR = contrast-to-noise ratio, DL = deep learning, DECT = dual-energy CT.}
    \vspace{-1em}
    \label{fig:5}
\end{figure}

\section{DISCUSSION}

\begin{table}[t]
\centering
\caption{Quantitative comparisons of HU accuracy of the original 140 images, the FCN denoised 140 kV images, and the DL-predicted 140 kV images.}
\resizebox{0.48\textwidth}{!}{\begin{tabular}{ccccc}
\toprule
         & Original & VMCT      &       & Improvement  \\
  Patient &  100 kV  &  (original  & VMCT  &  achieved with  \\
   ID   &  images   &   DECT) &  (DL-DECT) &  DL-DECT (\%) \\
\midrule
  Patient 1	&9.4	&11.2	&15.2&	35 \\
Patient 2	&10.7	&11.9	&15.9&	34 \\
Patient 3	&10.2	&10.7	&16.6&	55\\
\bottomrule
\vspace{0.1em}
\end{tabular}}\\
\label{tab:2}
\footnotesize{VMCT = virtual monochromatic CT, DECT = Dual-energy CT, DL = Deep learning.\\}
\end{table}

In this study, a strategy of using DL to generate VMCT images from a SECT scan is presented and the superior performance of the technique is demonstrated. Overall, the CNR of DL-based VMCT is prominently improved at an energy setting lower than 60 keV and higher than 110 keV as compared with the conventional VMCT imaging method. The maximum CNR of the DL-based VMCT imaging is higher than the that of conventional approach by 34\% to 55\% depending on the patients. More importantly, the maximum CNR of the proposed strategy is shifted from 80-85 keV for the conventional VMCT to 40 keV where the iodine has the highest contrast. This shift of maximum CNR is desirable feature and may be clinically useful in better visualizing low-contrast lesions such as hyperattenuating liver lesions.
Although the same low-energy CT are employed to reconstruct VMCT images, the DL-based VMCT has much lower noise. This noise reduction is attributed to the strong correlation of the noises in the low- and high-energy images of the DL-DECT images. The DL-predicted high-energy image is obtained by adding the residual image to the original low-energy image. Considering that the residual image has a very low noise level, the noise of the predicted image arises primarily from the original low-energy image. Thus the resulting DL-DECT images have similar noise textures as the input low-energy image. For this reason, the noise is reconciled during the reconstruction of the VMCT using DL-DECT images.
The DL-DECT predictive model generates the high-energy image based on both the HU values and the location of the pixels in the input image. For example, in the low-energy image, the bone marrow and the contrast-enhanced kidney may have similar HU values. But in the predicted 140 kV images, the HU value of kidney is lower than that of the bone marrow, suggesting the mapping procedure is not a linear or piece-wise linear function.
In addition to the virtual spectral monochromatic imaging, the DL-DECT images can also be applied to other DECT applications, such as differentiating intracerebral hemorrhage from iodinated contrast~\cite{gupta2010evaluation,tijssen2014role}, automated bone removal in CT angiography~\cite{sommer2009value,buerke2009dual}, virtual noncontrast-enhanced imaging~\cite{ferda2009assessment,graser2009dual,ho2012characterization,mangold2012virtual}, urinary stone characterization~\cite{boll2009renal,hidas2010determination,leng2015feasibility}. These applications should also benefit from the noise correlation of the DL-DECT images.
There are some limitations in this study. First, only one dual-energy technique (dual-source) and one energy setting (100kV/Sn 140 kV) were tested. For a different energy setting, a new model may need to be trained. Since different dual-energy techniques have different spectral characteristics, the mapping model is generally applicable only to a specific dual-energy technique. However, it is possible to train a general model using an assemble of images acquired from different dual-energy techniques or using images with carefully standardized calibration measurements. Second, this study is mainly focused on abdominal CT scan. Since higher iodine CNR of the VMCT images is achieved at lower energies where iodine has better contrast, the proposed approach can be applied to other DECT applications with improved visualization of low-contrast lesions. A quantitative clinical assessment along this line is out of the scope of this study but will be done in the future.

\section{Conclusion}

we have developed a DL-based virtual monochromatic spectral imaging technique. An important benefit of the proposed strategy is that it enables us to perform VMCT imaging without the need of a high-end DECT scanner. Furthermore, the characteristics of the image noise and iodine CNR of the VMCT images so obtained is very different from that reconstructed using conventional DECT images. The reduction in VMCT image noise and improvement in iodine CNR at low energy range may afford new opportunities for improved clinical practice.


%

%



%

\ifCLASSOPTIONcaptionsoff
  \newpage
\fi




%

\end{document}